\documentclass[a4paper,11pt]{article}
\usepackage{jinstpub} 
\usepackage{lineno}

\pdfoutput=1
\usepackage[utf8]{inputenc}
\usepackage{float}
\usepackage{lmodern}

\usepackage{newtxtext,newtxmath}
\usepackage{graphicx}
\usepackage{url}
\usepackage{hyperref}
\usepackage{color}
\usepackage{enumitem}
\usepackage{subfigure}
\usepackage[dvipsnames]{xcolor}
\hypersetup{allcolors=[rgb]{0.0 0.0 0.6},linkcolor=[rgb]{0.75 0.05 0.05}}
\usepackage{bm}
\usepackage{epsfig}

\usepackage{slashed}
\usepackage{color}
\usepackage{accents}
\usepackage{url}
\usepackage[dvipsnames]{xcolor}
\usepackage{cancel}
\usepackage[normalem]{ulem} 
\usepackage[colorinlistoftodos]{todonotes}
\usepackage{multirow}
\usepackage{natbib}
\usepackage{dsfont}
\usepackage{bbold}
\usepackage{xspace}
\usepackage{siunitx}
\DeclareSIUnit\sq{\ensuremath{\Box}} 

\newcommand{\um}{$\mu$m\xspace}
\newcommand{\uA}{$\mu$A\xspace}

\newcommand{\exclude}[1]{}

\begin{document}

\author[a]{Christina Wang,}
\author[a]{Cristi\'an Pe\~na,} 
\emailAdd{christiw@fnal.gov}
\author[b]{Adolf Bornheim,} 
\author[a]{Shuoxing Wu,}
\author[b]{Alexander Albert,}
\author[b]{Thomas Sievert,}

\author[a]{Artur Apresyan,} 
\author[c]{Emanuel Knehr,}
\author[c,d]{Boris Korzh,}
\author[c]{Jamie Luskin,}
\author[b]{Ludovico Mori,}
\author[b]{Sahil Patel,}
\author[b]{Guillermo Reales Guti\'errez,}
\author[d]{Manish Sahu,}
\author[c]{Ekkehart Schmidt,}
\author[c]{Matthew Shaw,}
\author[b]{Elise Sledge,}
\author[b]{Maria Spiropulu,} 
\author[d]{Towsif Taher,}
\author[a,b]{Si Xie}

\affiliation[a]{Fermi National Accelerator Laboratory, Batavia, IL 60510, U.S.A.}
\affiliation[b]{California Institute of Technology, Pasadena, CA 91125, U.S.A.}
\affiliation[c]{NASA Jet Propulsion Laboratory, Pasadena, CA 91011, USA}
\affiliation[d]{Group of Applied Physics, University of Geneva, CH-1205, Geneva, Switzerland}

\title{Towards High-Efficiency Particle Detection Using Superconducting Microwire Arrays}


\abstract{
We present a detailed study of an 8-channel $1\times1$~mm$^{2}$ WSi superconducting microwire single photon detector (SMSPD) array exposed to 120~GeV hadron beam and 120~GeV muon beam at the CERN Super Proton Synchrotron H6 beamline.
Following up on our first detailed characterization of the efficiency and response of an SMSPD fabricated on a 3~nm WSi film, we report measurements of enhanced particle detection efficiency using a sensor fabricated from a thicker 4.7~nm-thick WSi film. 
We also report the first SMSPD detection efficiency measurement made for muons. 
Measurements are enabled by a silicon tracking telescope providing 10~$\mu$m in-situ spatial resolution. 
The results show a fill factor-normalized detection efficiency of 75\% and a time resolution of about 130~ps across pixels.
These findings represent a significant advancement toward developing high-efficiency SMSPD charged particle tracking systems with simultaneous precision timing, with potential applications in future accelerator-based experiments such as the FCC-ee and Muon Collider.
}

\maketitle
{
  \hypersetup{linkcolor=black}
  \tableofcontents
}


\section{Introduction}
\label{sec:intro}

Superconducting Nanowire Single Photon Detectors (SNSPDs) are a leading detector technology for single-photon detection with diverse applications in optical communications~\cite{Mao:18,PhysRevLett.124.070501}, quantum information science~\cite{PRXQuantum.1.020317,Takesue:15, Shibata:14, PhysRevA.90.043804,Najafi_2015,Weston:16,Mueller:24}, and astronomy~\cite{PhysRevA.97.032329,PhysRevLett.123.070504, 10.1117/1.JATIS.7.1.011004}.
SNSPDs are exceptional in their ultra-low energy threshold of below 0.04~eV (or above 29$\mu$m)~\cite{Taylor:23}, dark counts as low as $10^{-5}$~Hz~\cite{7752769,Shibata:14, Chiles:2021gxk}, and pico-second level time resolution~\cite{Korzh:2020, Mueller:24}.
These properties make them highly attractive as next-generation detector technologies for accelerator-based high-energy physics experiments.

Despite these advantages, until recently, the use of SNSPDs in high energy physics (HEP) experiments has been limited due to their small active area (100~$\mu \text{m}^2$), being comprised of nanowires that are order of 100~nm wide.
Recent progress in the developments of thin superconducting films have, however, enabled the fabrication of large area (mm$^2$) detectors with micrometer-width superconducting wires~\cite{10.1063/5.0150282, 10.1063/5.0044057, 10.1117/1.JATIS.7.1.011004,Wang_2025, Korneeva2018, micron1, ChilesSMSPD2020,Charaev2020, Lita2021, XSMSPD, NbNMicrostrip, Protte2022}.
In this work, we refer to such devices as superconducting microwire single-photon detectors (SMSPDs). 
This advancement in large active area makes SMSPD a promising sensor for dark matter detection experiments~\cite{BREAD:2021tpx, Chiles:2021gxk} and a potential innovative detector technology for future accelerator-based experiments.

The first detailed measurement~\cite{Pena:2024etu} on the detection efficiency of SMSPD arrays to GeV energy particles was performed by our group with 120~GeV proton beam and 8~GeV electron and pion beam at the Fermilab Test Beam Facility (FTBF).
In the previous publication, we measured a 60\% fill factor-normalized detection efficiency and a 1.15~ns time resolution for the first time for an SMSPD array. The fill factor-normalized efficiency presents the efficiency of the active area of the SMSPD array, assuming particles incident on the non-superconducting areas do not produce a detectable signal.
The SMSPD tested was fabricated with 1.5~\um wide wires on a 3~nm WSi film sputtered from $\text{W}_\text{50}\text{Si}_\text{50}$.
In this work, we report a full characterization of the improved detection efficiency and time resolution of an 8-pixel $1\times1$ mm$^2$ SMSPD array fabricated with 1~\um wide wires on a thicker 4.7~nm WSi film sputtered from $\text{W}_\text{30}\text{Si}_\text{70}$. 
The measured fill factor-normalized detection efficiency is approximately 75\% and the time resolution was measured to be 130~ps.
The SMSPD array was characterized under the exposure of 120~GeV hadrons and 120~GeV muons from the CERN Super Proton Synchrotron (SPS) H6 beam line.

The SMSPD array under test is described in Section~\ref{sec:snspd}.
Section~\ref{sec:setup} describes the experimental setup at the beam line. 
The results of the full characterization of the SMSPD array are presented in Section~\ref{sec:results}.
Finally, the summary is presented in Section~\ref{sec:summary}.

\section{Superconducting microwire single photon detector array}
\label{sec:snspd}
\begin{figure}[ht]
	\centering
	\includegraphics[width=0.435\linewidth]{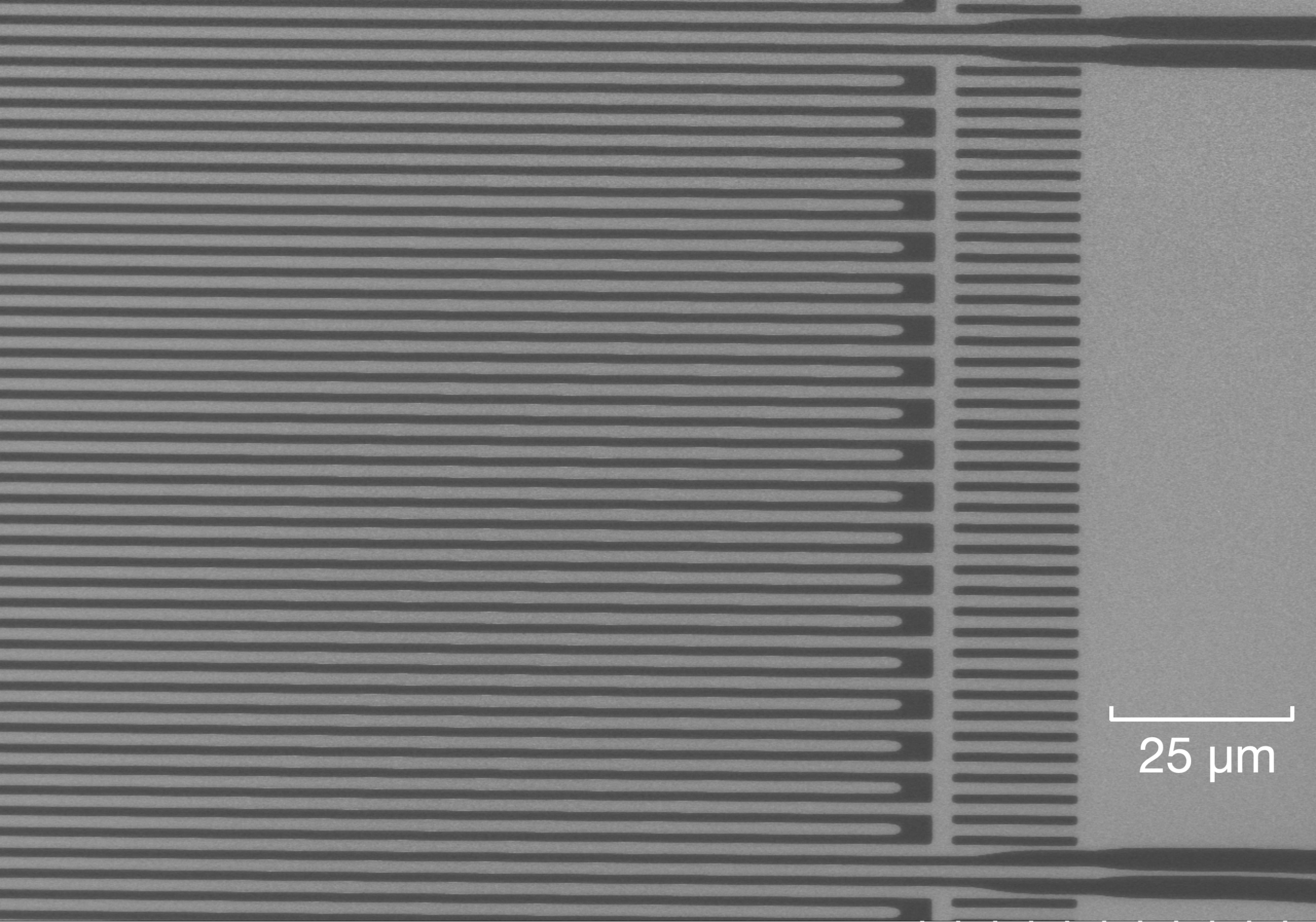}
	\includegraphics[width=0.45\linewidth]{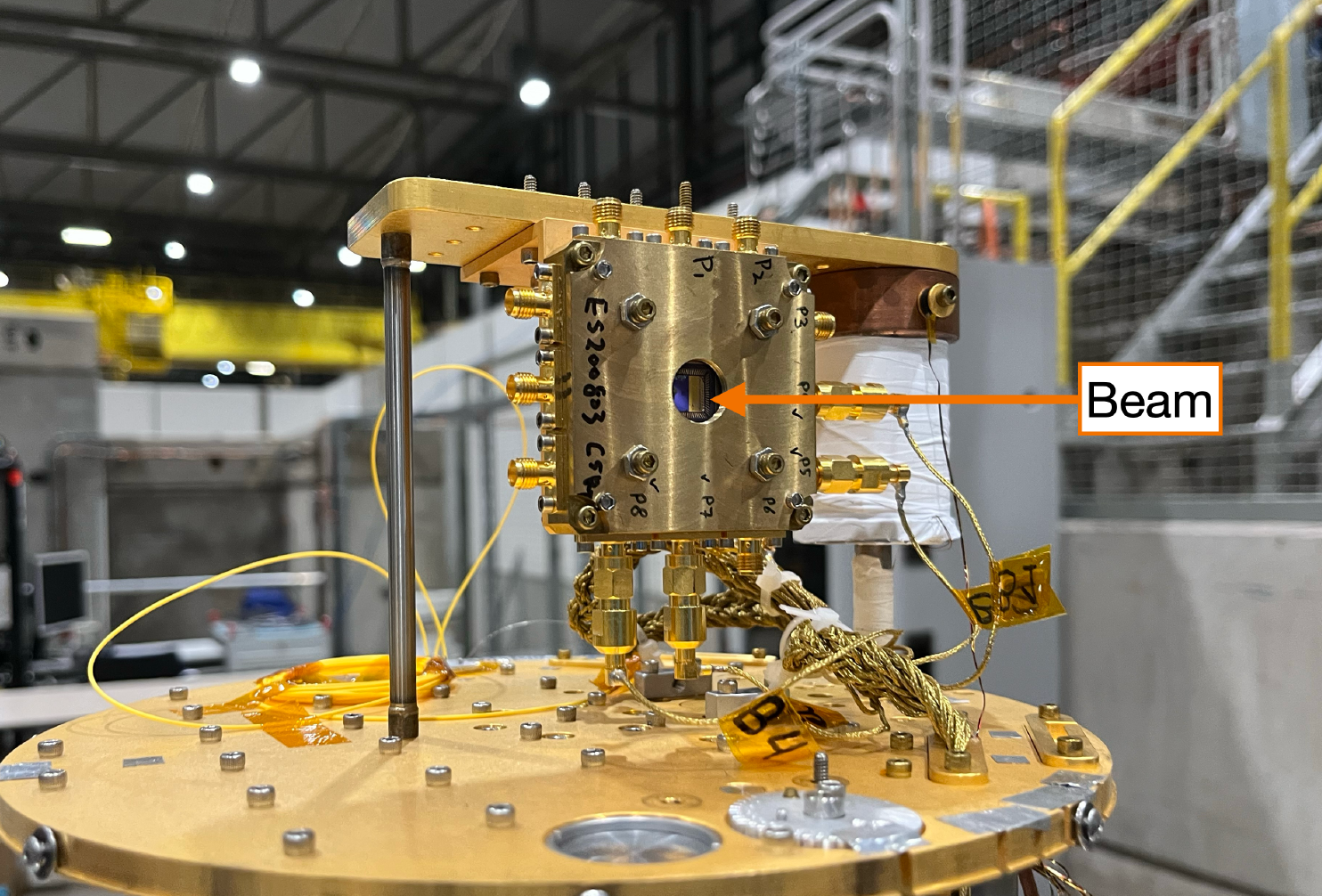}
 	\caption{A scanning electron micrograph of the meander structure of the sensor (left) and photograph of the SMSPD under study enclosed in a dark box attached to the cold plate in the cryostat (right).
}
	\label{fig:Mounting}
\end{figure}

The detector under test is a $1\times1$~mm$^{2}$ 8-channel SMSPD array fabricated from 4.7~nm thick WSi film. 
The fabrication was performed at the Jet Propulsion Laboratory.
Each pixel has a size of $0.125\times1$~mm$^{2}$.
The WSi film was sputtered from a $\text{W}_\text{30}\text{Si}_\text{70}$ target and deposited onto an oxidized silicon substrate with a 240~nm-thick oxide.
The final film stoichiometry was measured to be $\text{W}_\text{42}\text{Si}_\text{58}$.
The sheet resistance is measured to be 0.75~\si{k\ohm}/\si{\sq} at room temperature.
The critical temperature is 1.85~K.
The SMSPD array was patterned using electron beam lithography with 1~\um-wide wires meandering with a 3~\um gap width, amounting to a $25\%$ fill factor.
After etching, the microwires were covered with 10~nm of $\text{SiO}_\text{2}$ for passivation.
More details on the electron beam lithography of SMSPD can be found in Refs.~\cite{10.1063/5.0150282, 6994823}.
In this work, each pixel has an individual single-ended readout.
A schematic of the meander structure of the sensor is shown in Figure~\ref{fig:Mounting} (left).


The SMSPD is cooled down to 0.8~K with a PhotonSpot helium sorption fridge that has a hold time of about 30 hours.
The SMSPD was mounted vertically on a copper L-bracket from the base plate.
It was mechanically and thermally coupled to have normal incidence with respect to the beam, as shown in Figure~\ref{fig:Mounting} (right).

In this test beam campaign, we biased, amplified, and read out four of the available eight pixels independently.
A two-stage cryogenic DC-coupled amplifier operating at 40~K was developed by our group to provide a total gain of 30~dB for signal in the 10~MHz to 1~GHz frequency range, similar to that used in Ref.~\cite{Pena:2024etu,10.1063/5.0150282}, but more optimized for slower signals from large area SMSPDs.
The first stage of the amplifier is based on a low noise high-electron-mobility transistor and the second stage is based on a silicon germanium amplifier.
The DC-coupled amplifier also simultaneously provides SMSPD biasing through the same signal cable.
The signal from each SMSPD pixel is connected through RF cables to room temperature.
Subsequently, the RF signals are split into two, one of which is recorded by a high-rate time-to-digital converter (Swabian Time Tagger) and the waveform of the other is recorded by an oscilloscope, described in more detail in Section~\ref{sec:setup}. 
To measure the time resolution more accurately, a subset of the data were collected with additional external amplifiers at room temperature to improve the signal-to-noise ratio.

\begin{figure}[ht]
	\centering
	\includegraphics[width=0.6\linewidth]{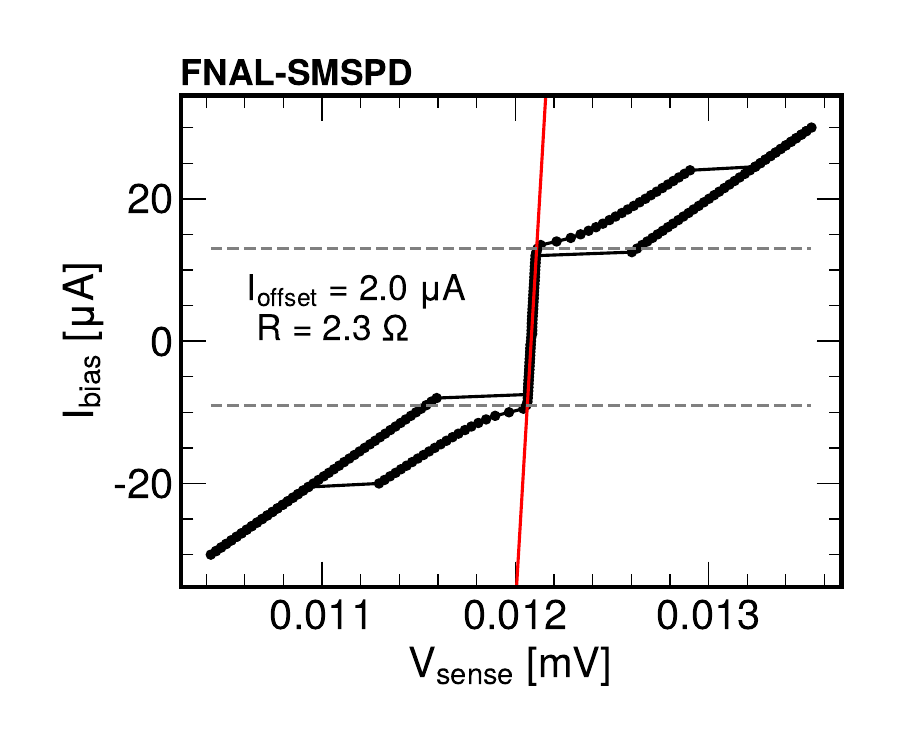}
 	\caption{The bias current ($I_\text{bias}$) with respect to voltage across one of the SMSPD pixels ($V_\text{sense}$). The linear region in the middle is the region when the SMSPD is superconducting. 
    }
	\label{fig:IVCurve}
\end{figure}

Since the bias current of the SMSPD is provided through the DC-coupled amplifier, it not only depends on the bias voltage provided, but also depends on the DC offset of the amplifier and the parasitic resistance of the SMSPD, both of which might vary between different cool downs.
Therefore, the current–voltage characteristic (IV curve) of the SMSPD pixels were measured at the beginning of every cool down cycle to derive the DC offset of the cryogenic amplifier and the parasitic resistance of the SMSPD and their variations.
An example IV curve is shown in Figure~\ref{fig:IVCurve}.
The parasitic resistance is extracted by fitting a linear function in the superconducting region. 
The DC offset is determined by the mean of the two bias currents (positive and negative) where the transition between superconducting and normal state occurs. 
The SNSPD bias current is then computed by dividing the bias voltage by the overall circuit resistance including the parasitic resistance, followed by the DC offset correction. 
Variations in the parasitic resistance of the SMSPD and the DC offset of the cryogenic amplifier across cool downs result in an uncertainty of 0.14~\uA in the SMSPD bias current, representing a few percent relative uncertainty.

\section{The experimental setup at the CERN SPS H6 beam line}
\label{sec:setup}

The data presented in this paper were collected at the CERN SPS H6 beamline, which provides secondary particle beams produced by the interaction of the primary proton beam with a momentum of 400~GeV from the SPS on a thin beryllium plate primary target.
The secondary particle beams are composed mainly of charged pions and protons, with a few percent of positrons and kaons~\cite{Atherton:133786}.
The primary proton beams are slowly extracted from the SPS towards the primary target, delivering about one to three 4.8~s-long spills every 40 seconds.
The number of spills depend on the number of facilities and users that require SPS extraction at the time.
In this paper, we will report on the results of the SMSPD response to a 120~GeV hadron beam, mainly composed of pions and protons, and a 120~GeV muon beam, selected by tuning the magnets and collimators downstream.

The H6 beamline is equipped with a beam telescope to measure the position of each incident particle precisely.
The telescope consists of six sensor planes using MIMOSA 26 monolithic active pixel devices with 18.4~\um pitch~\cite{Jansen:2016bkd}, five of which were in use during our testbeam campaign.
Since each SMSPD pixel is only 125~\um wide, it is important to optimize the spatial resolution of the telescope to be able to give an accurate measurement of the detection efficiency of the SMSPD.
To optimize the position resolution measurement from the telescope, the SMSPD is placed in between the telescope with three sensors upstream and two sensors downstream, where the modules are placed as close to the cryostat as possible. 
During this data-taking period, the spatial resolution of the telescope at the SMSPD was measured to be 10.4~\um$\pm$ 0.1~\um.
The spatial resolution of the telescope was characterized by studying the shape of the efficiency turn-on curve at the straight edge of an Low Gain Avalanche Detector (LGAD) that is placed immediately upstream of the cryostat.
Since the edge of the active area of the LGAD is much sharper than the 1--2~\um scale~\cite{Heller_2022}, any turn-on effect is attributed to the tracker resolution and can be extracted by a fit to an error function. 
The same method used on the edge of the SMSPD sensor also resulted in a measured resolution of 10~\um, indicating that the impact of potential fabrication or sensor response imperfections of the SMSPD is negligible. 

A Photek 240 micro-channel plate (MCP-PMT) detector operated at -3.8~kV is located downstream of the telescope and the SMSPD, and provides a precise reference timestamp with resolution below 10~ps~\cite{Ronzhin:2015idh}.

The SMSPD and MCP-PMT waveforms are acquired using a Lecroy Waverunner 8208HD oscilloscope. 
This oscilloscope features eight readout channels with a bandwidth of 2~GHz and a sampling rate of 10~GS/s per channel. 

The trigger signal originates in an independent device, a customized 4~mm$\times$5~mm LYSO crystal coupled to a silicon photomultiplier (SiPM) to limit the triggered events to a small area around the SMSPD.
The signal from the SiPM is sent to one of the oscilloscope channels and used as a trigger.
The trigger output of the oscilloscope is then sent to the trigger logic unit of the tracking telescope.

\begin{figure}[ht]
	\centering
	\includegraphics[width=0.95\linewidth]{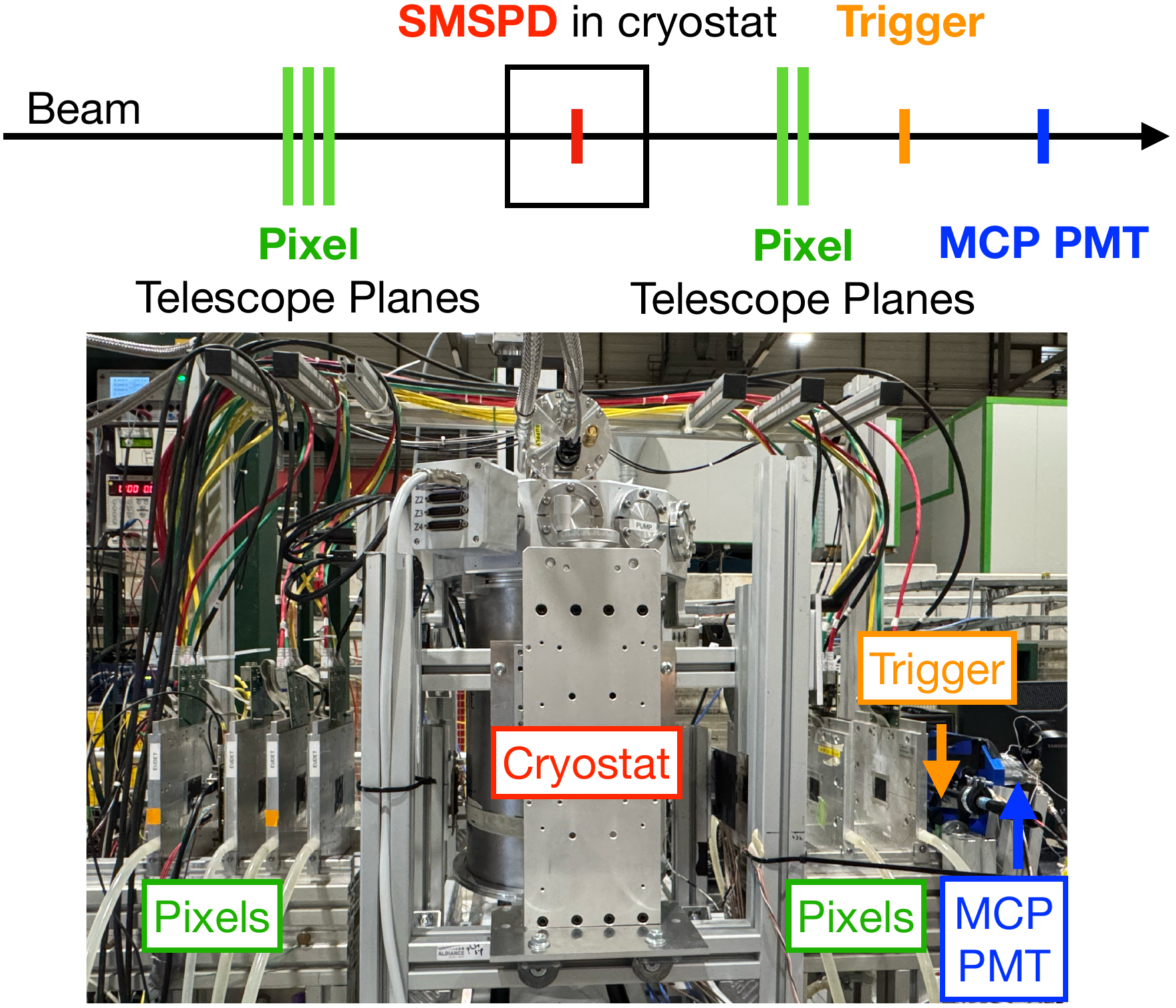}
 	\caption{
  A schematic diagram (top) and photograph (bottom) of the SMSPD under study and the reference instruments along the beamline. 
}
	\label{fig:FTBFSetup}
 
\end{figure}

Events are built by merging the telescope and oscilloscope data offline matching trigger counters from each system.
The telescope data are reconstructed with the Corryvreckan software~\cite{Dannheim:2020jlk}.
A schematic of the experimental setup is shown in Figure~\ref{fig:FTBFSetup} (top), which represents the arrangement of the cryostat hosting the SMSPD with respect to the telescope tracker, trigger, and MCP-PMT. 
Figure~\ref{fig:FTBFSetup} (bottom) shows a photograph of the setup.

\section{Experimental Results}
\label{sec:results}

In this section, we present detailed study of the SMSPD response to 120~GeV hadrons and muons, including results on the sensor signal properties, time resolution, and detection efficiency. 

Events considered in our analysis are required to have a high-quality track with a measured hit position inside the SMSPD active area.
A high-quality proton track must have hits in all 5 MIMOSA planes and a reduced $\chi^2$ less than 5.
In addition, to identify tracks unambiguously, we require there to be exactly one reconstructed track per event.
To ensure a reliable reference for particle time, the event must also have an MCP-PMT signal consistent with a minimum ionizing particle (MIP).

\begin{figure}[ht]
	\centering
        \includegraphics[width=0.45\linewidth]{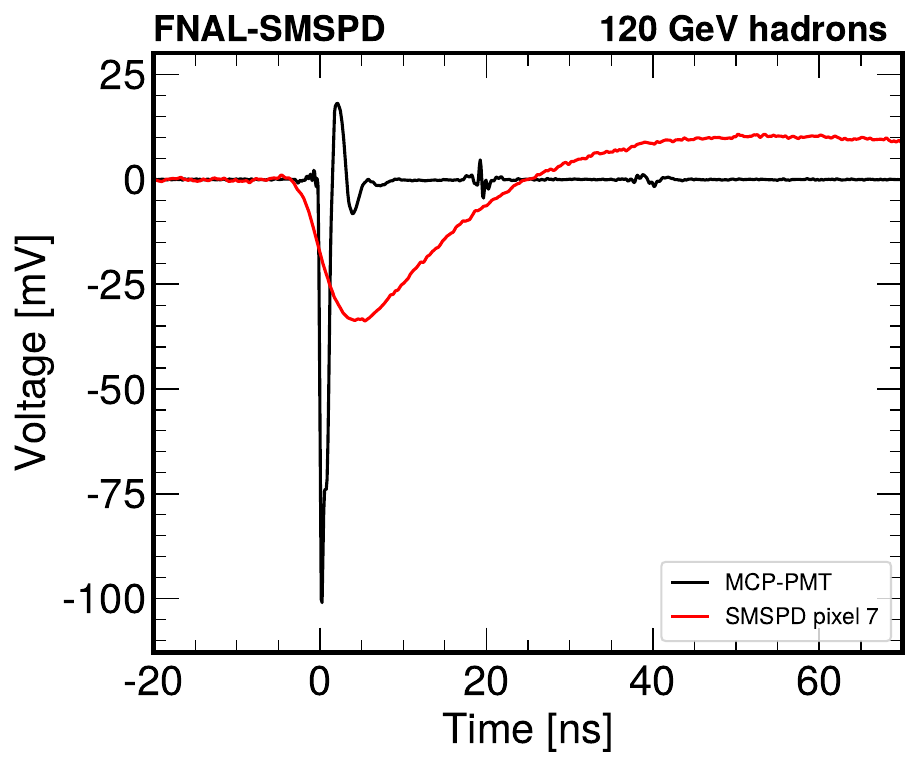}
	\includegraphics[width=0.45\linewidth]{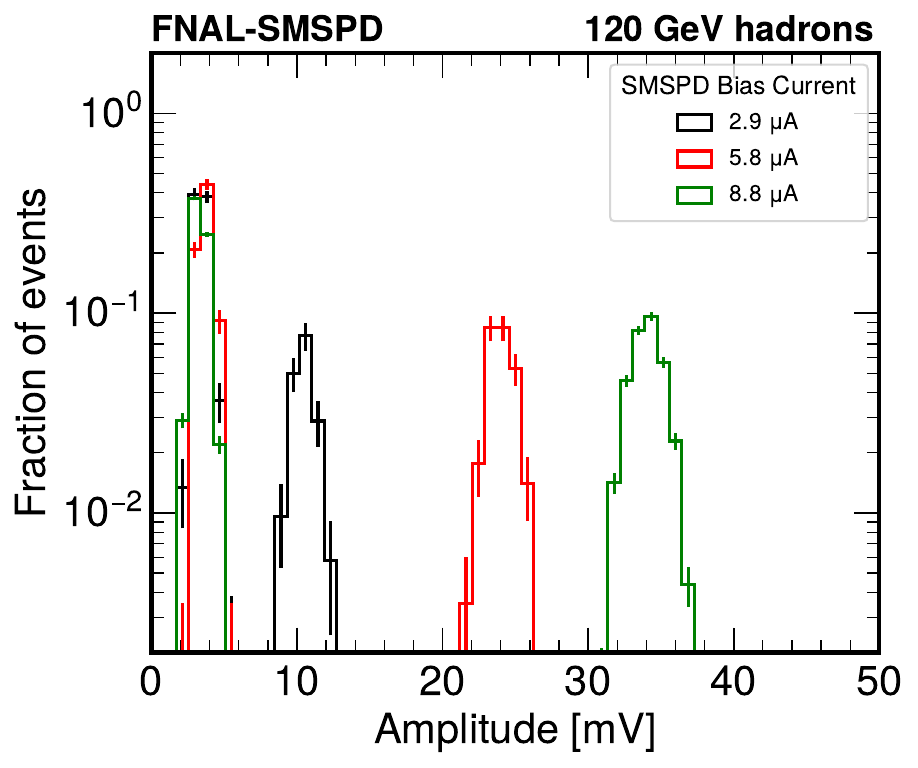}
 	\caption{The average pulse shapes (left) from the MCP-PMT and pixel 7 of the SMSPD from triggered events are shown. 
    The reconstructed amplitude distributions (right) from the pulse shapes of SMSPD pixel 7 for various bias currents are shown, where the peaks on the left are from baseline electronic noise and the peaks on the right are from signal pulses. 
   The mean of the signal amplitude increases linearly with the bias current as expected.
}
	\label{fig:SignalWaveforms}
\end{figure}

We first present our result for hadron beams.
Averaged waveforms for triggered hadron events produced by one of the SMSPD pixels in coincidence with the MCP-PMT are shown in Figure~\ref{fig:SignalWaveforms} (left). 
The waveforms for each pixel have similar shapes, while the reconstructed amplitude of the waveforms increases linearly as the bias current increases, as shown in Figure~\ref{fig:SignalWaveforms} (right). 
In order to efficiently select signal events for every operating bias current, the amplitude threshold for each bias current is set to be at least 3 standard deviations below the mean of the signal peak.

\begin{figure}[htb!]
	\centering
	\includegraphics[width=0.49\linewidth]{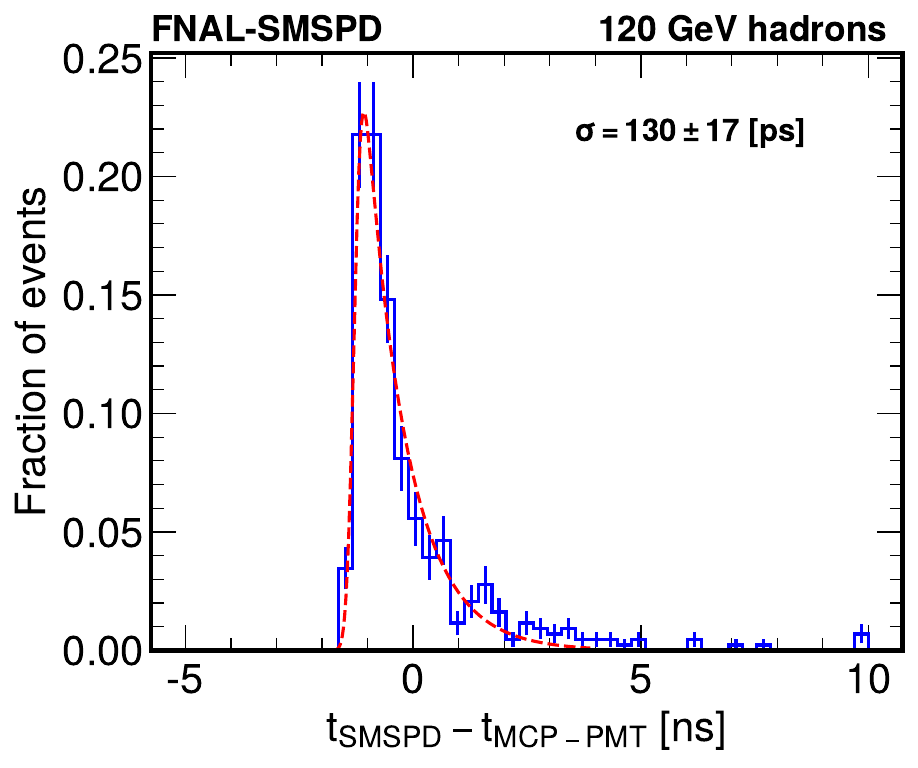}
 	\caption{The time difference between the time of arrival of SMSPD pixel 7 signal (t$_\mathrm{SMSPD}$) and MCP-PMT signal (t$_\mathrm{MCP-PMT}$) is shown, demonstrating a $130\pm17$~ps time resolution ($\sigma$) of the SMSPD.
  }
	\label{fig:TimeResolution} 
\end{figure}

The time stamps of the SMSPD and MCP-PMT signal are both determined by performing a fit to the rising edge of the pulse to extract the time at which the pulse reaches 50\% of the maximum amplitude.
The time difference between the MCP-PMT and one of the SMSPD pixels is shown in Figure~\ref{fig:TimeResolution} and observed to be similar across different SMSPD pixels and different values of SMSPD bias current.
The time resolution is determined by fitting an exponentially modified Gaussian (EMG), which fits the distribution more closely than a normal distribution and was thus adopted in previous studies~\cite{PhysRevB.96.184504,Caloz:2017gbz,Korzh:2020,Wang_2025}.
Since the MCP-PMT has been measured to have a time resolution of less than 10~ps~\cite{Ronzhin:2015idh}, the fitted width of the distribution is dominated by the time resolution of the SMSPD, which is measured to be $130\pm17$~ps.
Due to the non-negligible meander lengths of superconducting wires, we expect a geometric jitter effect arising from the particles detected at different positions along the wires~\cite{Korzh:2020,Kuzmin:2018pbk,Caloz:2017gbz}.
This result is consistent with the 138~ps timing resolution previously measured for 1064~nm photons using an SMSPD array with the same film stoichiometry and meandering geometry~\cite{10.1063/5.0150282}.
In the future, we plan to improve the time jitter by engineering devices with faster rise times and optimizing the readout scheme to correct for the jitter due to the varying hit position along the wire.

In order to measure the device efficiency, events are required to have a well measured track in the MIMOSA telescope, with a measured hit position inside the SMSPD sensor active area. 
Events considered in the efficiency measurement are also required to have a time stamp within 5~ns from the time stamp of a MIP-like signal measured in the MCP-PMT. 
The detection efficiency is measured as the fraction of these events which also have a signal above threshold in an SMSPD pixel.
The measurement is performed as a function of the incident particle position. 

\begin{figure}[ht]
	\centering
        \includegraphics[width=0.47\linewidth]{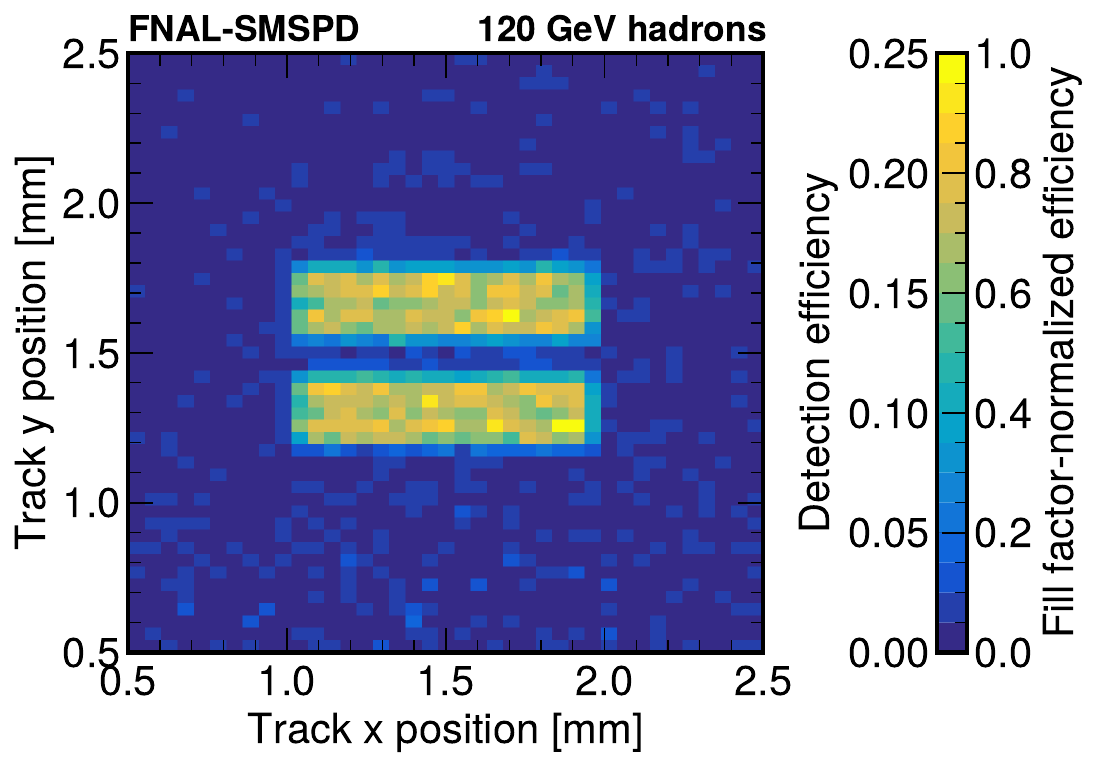}
    \hspace{0.03\linewidth} 
 	\includegraphics[width=0.40\linewidth]{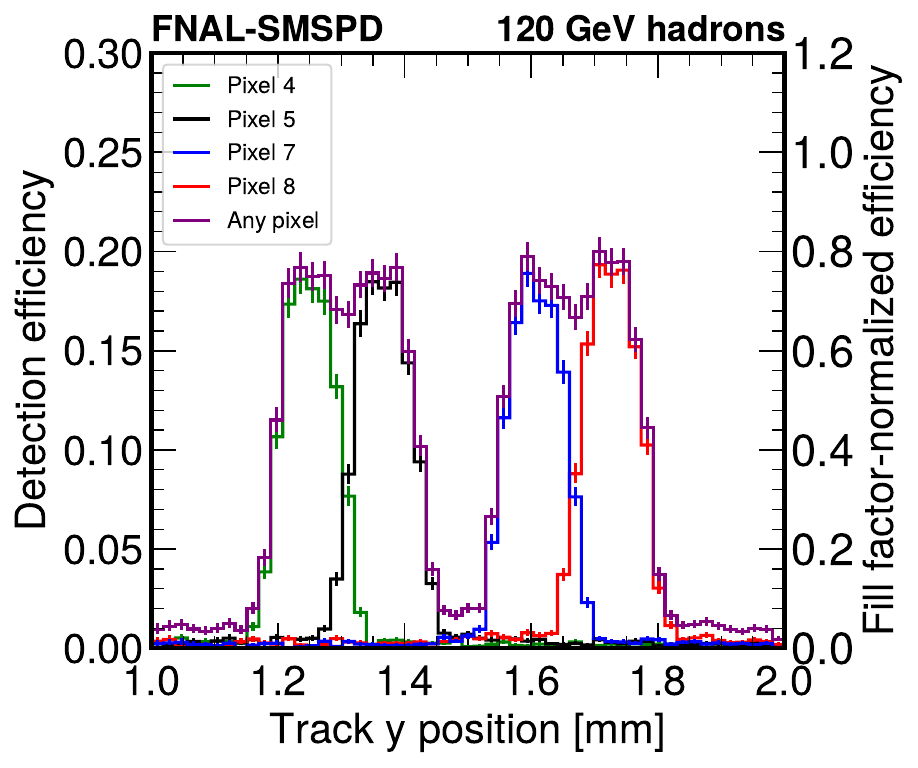}
  
 	\caption{The (fill factor-normalized) detection efficiency of the four SMSPD pixels are shown. 
  The efficiency as a function of the incident track position in the x and y directions (left) and in the y direction only (right) while integrating along the x-axis a 0.7~mm length from the center of each SMSPD pixel. 
  The data in both plots were recorded with the pixels 4, 5, 7, and 8 operating at 8.7, 9.1, 8.8, and 9.1~\uA, respectively.  
  The gap observed between pixel 5 and pixel 7 is expected and due to the fact that pixel 6 in between was not read out.
  }
	\label{fig:SNSPDEfficiency}
 
\end{figure}

Figure~\ref{fig:SNSPDEfficiency} (left) shows the hadron detection efficiency and fill factor-normalized efficiency when a signal is present in any of the four SMSPD pixels as a function of the measured track x and y positions.
The fill factor-normalized efficiency, as defined in Section~\ref{sec:intro}, is the efficiency of the active area of the SMSPD array, assuming particles incident on the non-superconducting areas do not produce a detectable signal.
The gap observed between pixel 5 and pixel 7 is expected and due to the fact that pixel 6 in between was not read out.
Figure~\ref{fig:SNSPDEfficiency} (right) shows the (fill factor-normalized) detection efficiency as a function of the track y position, integrating along the x-axis a 0.7~mm length from the center of each SMSPD pixel.
The integration area is slightly smaller than the full 1~mm length, to ensure that the detection efficiency is minimally impacted by the imperfect spatial resolution of the reconstructed tracks at the edges of the sensor.
Clear contribution can be seen from each individual pixel, showing high fill factor-normalized detection efficiency of about 75\% and good uniformity across the pixels.

\begin{figure}[ht]
	\centering
  \includegraphics[width=0.45\linewidth]{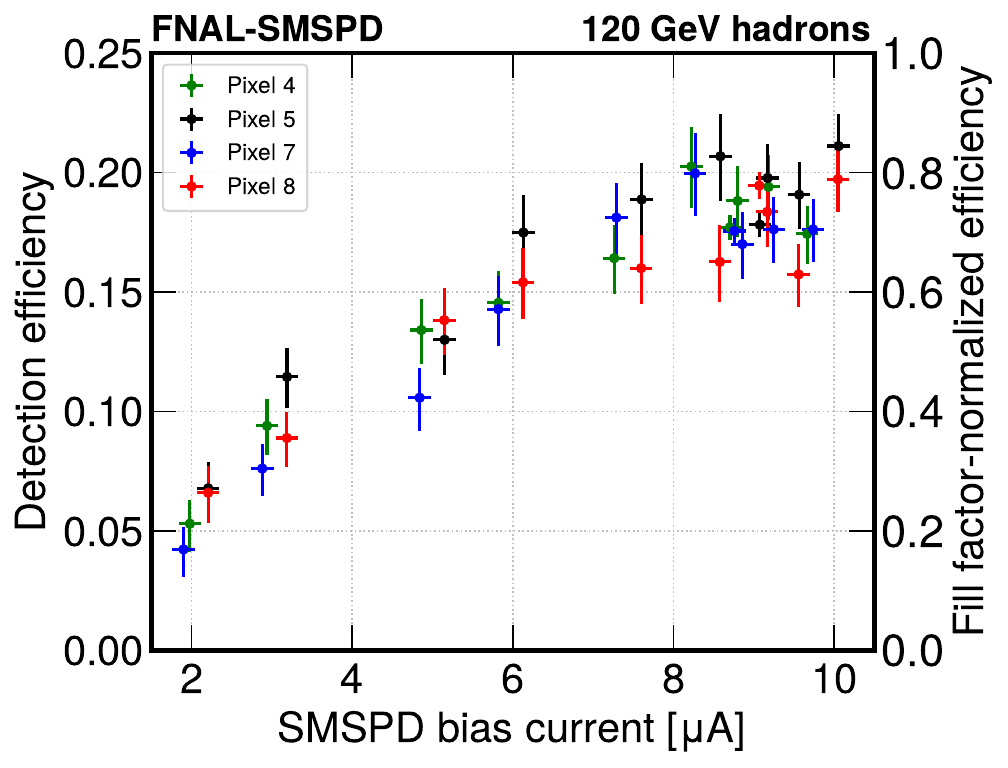}
      \hspace{0.05\linewidth} 
  \includegraphics[width=0.45\linewidth]{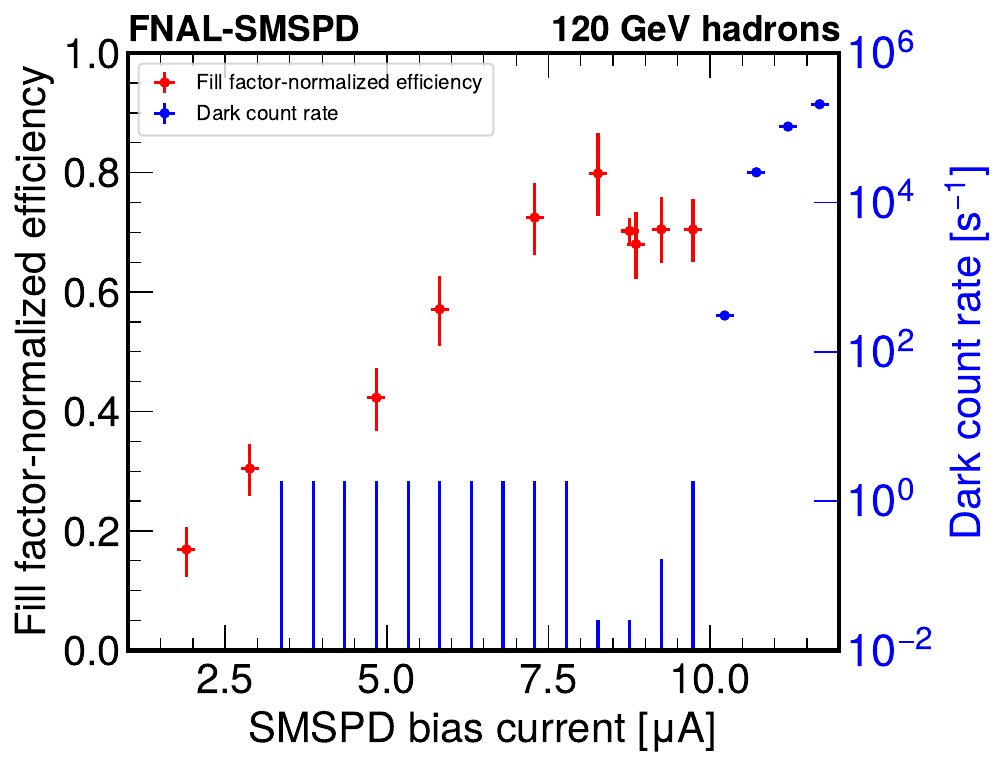}
 	\caption{The (fill factor-normalized) detection efficiency of all four measured SMSPD pixels (left) and the fill factor-normalized detection efficiency and DCR of pixel 7 (right) are shown.
  The efficiency measurements include the associated statistical uncertainty.
  The vertical error bars on the dark count rate correspond to statistical uncertainties, which is implemented with the Garwood method~\cite{10.1093/biomet/28.3-4.437} using a 68\% confidence interval which provides coverage for event counts following Poisson distributions, especially when statistics is low.
  The 0.14~$\mu$A uncertainty on the SNSPD bias current are from variations of the DC offset in the DC-coupled cryogenic amplifier and the parasitic resistance of the SMSPD across different cool down cycles.
    For bias currents below 10~\uA, zero dark counts were measured given the integration time, so only the error bars are shown.
    The varying statistical uncertainties across different measurements are due to differences in integration time.
    }	\label{fig:SNSPDEfficiencyVsBias_proton}

\end{figure}

Finally, the inclusive (fill factor-normalized) detection efficiency of each of the four pixels, calculated by integrating over a fiducial region of $700\times60~\mu\text{m}^2$ for each SMSPD pixel are shown in Figure~\ref{fig:SNSPDEfficiencyVsBias_proton} (left).
The fiducial region is smaller than the active area of the SMSPD to ensure that the inclusive efficiency calculation is not biased by the non-negligible telescope spatial resolution.
Similar and consistent efficiencies are observed across the four readout pixels.
An uncertainty on the bias current due to the variations of the DC offset in the DC-coupled cryogenic amplifier and the parasitic resistance of the SMSPD across different cool down cycles, as described in Section~\ref{sec:snspd}, is accounted for and estimated to be 0.14~$\mu$A.
The fill factor-normalized detection efficiency of one of the SMSPD pixels is overlayed with its dark count rate (DCR), as shown in Figure~\ref{fig:SNSPDEfficiencyVsBias_proton} (right).
The DCR is measured in situ with a Swabian Time Tagger immediately before the beam was turned on.
As shown in the figure, there exists an operating regime with bias current from 7.5--10~$\mu$A, where the SMSPD fill factor-normalized efficiency is above 70\% and DCR is below 1~Hz.

\begin{figure}[htb!]
	\centering
	\includegraphics[width=0.48\linewidth]{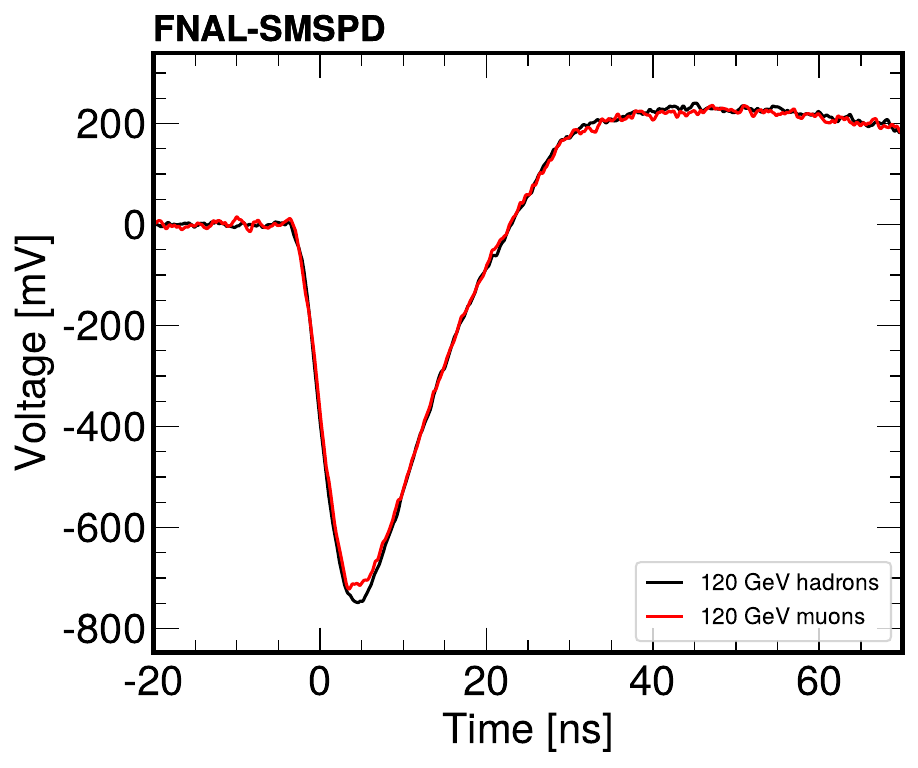}
	\includegraphics[width=0.48\linewidth]{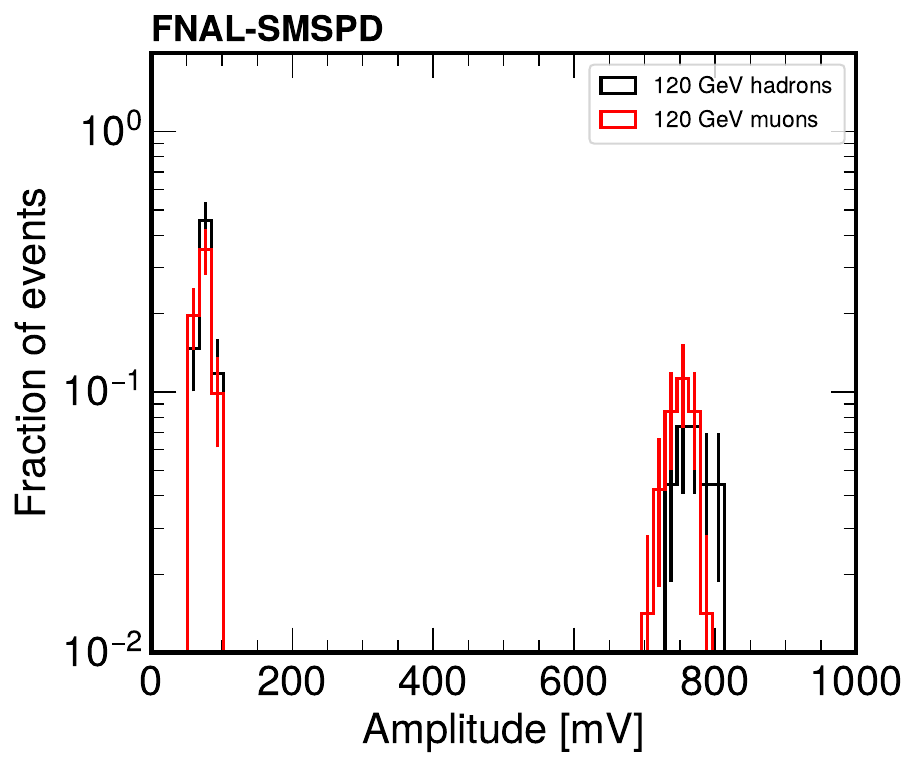}
 	\caption{The average pulse shapes (left) from the pixel 7 of the SMSPD from triggered hadron and muon events are shown.
    The data shown here includes an external amplifier to enhance the signal-to-noise ratio, thus the amplitude is much larger than that in Figure~\ref{fig:SignalWaveforms}.
    The reconstructed amplitude distributions (right) from the pulse shapes of SMSPD pixel 7 for proton and muon events are shown, where the peaks on the left are from baseline electronic noise and the peaks on the right are from signal pulses.
    The data in both plots were recorded with pixel 7 operating at 8.8~\uA. 
}
	\label{fig:SignalWaveformsMuons}
\end{figure}

Finally, we will briefly discuss our results for 120~GeV muon beams.
Events considered in the analysis for muons have the same requirement as the analysis for hadrons.
Events are required to have exactly one high-quality track with a position measurement inside of the SMSPD active area and a MCP-PMT signal corresponding to a MIP.
The SMSPD waveforms and amplitude distributions from muons are similar to that of hadrons, as shown in Figure~\ref{fig:SignalWaveformsMuons}. 
Therefore, the same amplitude threshold is used in the data analysis of muon beams.
For the muon beam data, we collected data at one bias current, similar to that in Figure 7, where the pixels 4, 5, 7, and 8 were operating at 8.7, 9.1, 8.8, and 9.1~\uA, respectively.  
The inclusive detection efficiency per pixel for muons, calculated by integrating the same fiducial region of each SMSPD pixel, is measured to be consistent with that of hadrons.

\section{Summary}
\label{sec:summary}


In summary, this paper presents detailed studies of an 8-channel $1\times1$~mm$^{2}$ WSi SMSPD array exposed to 120~GeV hadron beam and 120~GeV muon beam at the CERN SPS H6 beam line.
The absolute and calibrated detection efficiency was measured in detail for 120~GeV hadrons and for the first time for 120~GeV muons, enabled by the pixel tracker that provided precise spatial resolution of 10~\um.
The result demonstrated a consistently improved fill factor-normalized detection efficiency of 75\% across pixels and different particle types using a sensor fabricated from a thicker 4.7~nm-thick WSi film as compared to the 3~nm WSi film used in the previous study~\cite{Pena:2024etu}.
A possible explanation for the observed improvement in detection efficiency is the increased expected energy deposition in thicker films. 
We are actively conducting further studies to improve our understanding, and these results will be reported in a future publication.
Time resolution of 130~ps was measured using a MCP-PMT which provided ps-level reference time stamp.
The results presented in this paper represents a significant advancement towards developing high-efficiency superconducting wire detector systems for particle detection and identification at the next-generation accelerator-based experiments.

\acknowledgments
This manuscript has been authored by FermiForward Discovery Group, LLC under Contract No. 89243024CSC000002 with the U.S. Department of Energy, Office of Science, Office of High Energy Physics.
A.A., A.B., E.S., G.G., L.M., M.S., S.X., T.S are partially supported by the U.S. Department of Energy, Office of Science, Office of High Energy Physics, under Award No. DE-SC0011925.
A.A., C.P., C.W., S.W., S.X. are partially supported by the U.S. Department of Energy, Office of Science Accelerate Initiative Program Award under FWP FNAL 23-30. 
This work was partially supported by the Fermilab Laboratory Directed Research and Development (LDRD) program and New Initiatives Program. 
The University of Geneva acknowledges support from the SERI CHEF program (CH Experimental research at the FCC).
Part of this research was performed at the Jet Propulsion Laboratory, California Institute of Technology, under contract with the National Aeronautics and Space Administration.
We thank J.Allmaras and E.Wollman for the productive discussions on data analysis and interpretation.
We thank CERN for providing the test beam for the excellent performance of the accelerator and support of the test beam facility, in particular M. Doser, M. Jaekel, L. Zwalinski and M. Van Dijk. We also thank A. Rummler for help with the telescope tracker system and providing late night supports.

\bibliography{refs}
\bibliographystyle{jhep}
\end{document}